\documentclass[aps,pra,twocolumn,showpacs,superscriptaddress,groupedaddress]{revtex4}
\usepackage{graphicx} % Required for inserting images
\usepackage{amsmath}

\begin{document}

\title{Oscillation-Induced Frequency Generation in 1D Quantum Droplets under Harmonic–Gaussian Confinements}
\author{Jagnyaseni Jogania}
\affiliation{C. V. Raman Global University, Bhubaneswar, Odisha 752 054, India}
\author{Saurab Das}
\affiliation{Indian Institute of Information Technology Vadodara, Gujarat, India 382 028}
\author{Ajay Nath}
\affiliation{Indian Institute of Information Technology Vadodara, Gujarat, India 382 028}
\author{Jayanta Bera}
\affiliation{C. V. Raman Global University, Bhubaneswar, Odisha 752 054, India}

%\author{Suranjana Ghosh}
%\affiliation{Indian Institute of Technology Patna, Bihar, India 801106}
%\author{Utpal Roy}
%\affiliation{Indian Institute of Technology Patna, Bihar, India 801106}

\begin{abstract}
We explore the dynamical behavior of one-dimensional quantum droplets (QDs) governed by the extended Gross-Pitaevskii equation, under harmonic confinement supplemented by a static or time-dependent Gaussian spike (Gs) potential. Employing both variational analytical techniques and numerical simulations, we investigate the evolution of the root-mean-square (RMS) size, excitation spectrum, and phase-space dynamics of QDs. Our study reveals that while the harmonic trap sets the primary confinement, the Gs potential enables precise frequency tuning and control over droplet oscillations. A static Gs amplitude modifies the fundamental oscillation frequency depending on its sign, while a time-modulated Gs induces nonlinear dynamics, including higher harmonics and frequency mixing. Our analysis reveals that the resulting frequency spectrum is strongly influenced by inter- and intra-species interactions as well as by the parameters of the external trap. Notably, we establish a relationship between the frequency shift and the amplitude of the Gaussian spike. Wigner phase-space analysis further uncovers coherent rotational behavior, offering insights into hidden phase dynamics not apparent in real-space density profiles. These results highlight the role of time-dependent external potentials in tailoring the excitation landscape of quantum droplets, with implications for quantum simulation, precision control in atomtronics, and probing coherence in low-dimensional quantum fluids.

\end{abstract}

\maketitle

\section{Introduction}
Recent theoretical advances in the study of Bose-Einstein condensates have led to the identification of a novel class of self-bound quantum fluids known as quantum droplets (QDs)—ultradilute, liquid-like states that emerge in ultracold atomic gases and display characteristics distinct from those of classical fluids \cite{Malomed1,ttcher,Luo}. The formation of such droplets in weakly interacting Bose-Bose mixtures is a striking manifestation of beyond-mean-field (BMF) physics, where quantum fluctuations—captured by the Lee-Huang-Yang (LHY) correction—stabilize the system by counteracting the attractive mean-field interactions between different atomic components \cite{Petrov,Petrov1}. QDs are self-bound states stabilized purely by interparticle interactions, without requiring external confinement. Initially predicted in three-dimensional Bose mixtures by Petrov \cite{Petrov1}, QDs have since been realized experimentally in both two-component BECs \cite{Cabrera, Cheiney, Semeghini} and single-component dipolar condensates, where anisotropic dipole-dipole interactions combined with LHY quantum fluctuations support self-bound behavior \cite{Ferrier, Schmitt, Hertkorn}. These advances have enabled precise preparation, control, and imaging of QDs \cite{Barbut, Wenzel, Ferioli}, renewing interest in their rich non-perturbative many-body physics \cite{Lavoine,Maitri,Tylutki,Hu,Edmonds,Bhatia,Zhang,Das,Xia,Parisi,Astrakharchik,Cui,Sekino,Maitri2}. QDs provide a rich platform for exploring emergent phenomena such as supersolidity~\cite{Parit, Mukherjee}, dimensional crossovers beyond mean-field theory \cite{Lavoine, Zin1}, pattern formation \cite{Maitri3}, enhanced mobility~\cite{Kartashov}, reflectionless scattering \cite{Xia}, time crystal generation \cite{Nath1}, and the formation of droplet molecules via inter-droplet binding~\cite{Ferioli}. We have previously demonstrated higher harmonic generation in QDs subjected to tilted and driven quasi-periodic confinements \cite{Maitri3}, and in ultracold dipolar condensates \cite{bera2023}. Nonetheless, a thorough theoretical and numerical investigation of QD oscillation spectra—specifically under composite confinement formed by harmonic and time-dependent Gaussian spike (Gs) potentials—remains underexplored. In particular, the generation and control of alternating frequency spectra and their phase-space dynamics, especially within frameworks such as variational analysis and full numerical simulation, are largely absent from the current literature \cite{Jacob, Akram}.

In this work, we investigate the dynamics of 1D QDs confined in a composite potential composed of a harmonic trap and a static or time-dependent Gaussian spike (Gs) potential. The system is modeled using the 1D extended Gross-Pitaevskii equation (eGPE), which includes beyond-mean-field effects. We employ both a variational analytical method and numerical simulations using the split-step Fourier transform to analyze the evolution of the droplet’s root-mean-square (RMS) size, excitation spectrum, and phase-space behavior. Our findings show that while the harmonic trap provides the dominant confinement, the Gs potential offers precise control over the droplet dynamics. Varying the Gs amplitude from repulsive to attractive leads to a systematic increase in oscillation frequencies, enabling fine-tuning of excitation spectra. When the Gs is periodically modulated, the system exhibits nonlinear features such as higher harmonics and frequency mixing, indicative of driven many-body dynamics. A Wigner phase-space analysis reveals coherent rotational patterns, uncovering phase information not evident in the density evolution alone. Analytical results are obtained via a variational approach \cite{Otajonov, Chen, Saha}, while numerical simulations are performed using the split-step Fourier transform method. Our findings demonstrate excellent agreement between the analytical predictions and numerical results, validating the robustness of the proposed framework.

This work is structured as follows. In Sec. II, we present the theoretical framework for modeling QDs and develop an analytical formulation based on the variational approach. Section III applies this model to examine the modification of QD states, RMS size dynamics, and the resulting frequency spectrum under the influence of composite trapping potentials. In Sec. IV, we further analyze the RMS size oscillations and frequency characteristics of QDs in the presence of time-dependent perturbations. Section V provides a detailed numerical simulation of QD dynamics under temporally varying potential depths using the split-step Fourier method, allowing for a direct comparison with analytical predictions. Additionally, a Wigner phase-space analysis is carried out to gain further insight into the system's dynamical behavior. Finally, Sec. VI summarizes the main findings and outlines potential directions for future research.

\section{Theoretical model and dynamical equations}
In this model, we consider a homonuclear binary Bose–Einstein condensate mixture comprising two distinct hyperfine states of $^{39}$K \cite{Semeghini}. To simplify the analysis, we assume a symmetric binary configuration with an equal number of atoms and identical masses: $\psi_{1} = \psi_{2} = c_{0} \psi$, $N_{1} = N_{2} = N$, and $m_{1} = m_{2} = m$. The intra-species interaction strengths are taken to be equal, $g_{\uparrow\uparrow} = g_{\downarrow\downarrow} \equiv g (>0)$, defined as $2 \hbar^{2} a_{s} / (m a_{\perp}^{2})$. The inter-species interaction is taken to be attractive, $g_{\uparrow\downarrow} < 0$, leading to a quantum droplet regime when the effective interaction parameter $\delta g = g_{\uparrow\downarrow} + g > 0$ \cite{Astrakharchik}.

Under these assumptions, the binary BEC mixture can be effectively reduced to a single-component, dimensionless, one-dimensional extended Gross–Pitaevskii equation (eGPE) with first-order LHY quantum corrections \cite{Petrov, Astrakharchik, Nie}:

\begin{eqnarray}\label{eq:QD1}
i\frac{ \partial \psi}{\partial t} = - \frac{1}{2} \frac{ \partial^2 \psi}{\partial x^2} - \gamma_1 |\psi|\psi + \gamma_2 |\psi|^2\psi + V(x,t) \psi.
\end{eqnarray}

Here, the wavefunction, length, and time are expressed in units of $(2 \sqrt{g})^{3/2}/[\pi \xi(2|\delta g|)^{3/4}]$, $\xi$, and $\hbar^{2}/(m \xi^{2})$, respectively, where $\xi = \pi \hbar^{2} \sqrt{|g|}/(m g \sqrt{2})$ is the healing length of the system \cite{Astrakharchik}. The coefficients $\gamma_1 = (\sqrt{2m}/\pi \hbar) g^{3/2} = (\sqrt{2m}/\pi \hbar) [\sqrt{g_{11}g_{22}}]^{3/2}$ and $\gamma_2 = g_{12} + \sqrt{g_{11}g_{22}}$, with $g_{ii} = 4\pi a_{ii}/m_i$ and $g_{12} = 2\pi a_{12}/m_r$, $m_r = m_1 m_2 / (m_1 + m_2)$, characterize the beyond-mean-field (BMF) and effective mean-field (EMF) nonlinearities, respectively.

The quadratic nonlinearity arises from the LHY correction, contributing an effective attractive interaction, while the cubic term represents the repulsive mean-field contribution. Together, these nonlinearities facilitate the formation of self-bound quantum droplets.

The external potential consists of a harmonic trap superimposed with a time-dependent Gaussian spike at the center, and is given by \cite{Jacob, Akram}:

\begin{eqnarray}\label{eq:QD3}
V_{\text{ext}}(x,t) = \frac{\gamma^2 x^2}{2} + C(t)\exp\left(-\frac{x^2}{2\sigma^2}\right),
\end{eqnarray}

where $C(t)$ is the temporally modulated amplitude of the Gaussian spike:

\begin{equation}
C(t) = C_0 \left(1 + \alpha \cos(qt)\right).
\end{equation}

with $C_{0}$, $\alpha$, and $q$ are integers. The dynamical behavior of the condensate is analyzed using the Lagrangian variational (LV) approach under standard approximations \cite{Otajonov, Chen, Saha}. We consider a Gaussian trial function of the form:

\begin{eqnarray}
\psi(x,t) = A(t)\exp\left[-\frac{x^2}{2\omega(t)^2} + i\beta(t)x^2 + i\phi(t)\right], \label{ansatz}
\end{eqnarray}

where $\omega(t)$ denotes the condensate width, $\beta(t)$ the chirp parameter, and $\phi(t)$ the global phase. The total Lagrangian is obtained by integrating the Lagrangian density over space. The detailed steps involve computing individual contributions $L_1$ through $L_5$ from the terms of kinetic, potential, mean field, and quantum fluctuation, are outlined in Appendix A.

Applying the Euler-Lagrange equation,

\begin{eqnarray}
\frac{d}{dt}\frac{\partial L_{\text{eff}}}{\partial \dot{u}} = \frac{\partial L_{\text{eff}}}{\partial u}, \label{el}
\end{eqnarray}

for $u \in {\omega(t), \beta(t), \phi(t)}$, we derive the following equation of motion for the width $\omega(t)$:

\begin{eqnarray}
\omega''(t) + \gamma^2 \omega(t) - \frac{1}{\omega(t)^3} - \frac{2\sqrt{2}\sigma C(t)\omega(t)}{(2\sigma^2 + \omega(t)^2)^{3/2}} \nonumber\\
\frac{\sqrt{N_A/(2\pi)}}{\omega(t)^2} + \frac{2\sqrt{6N_A}}{9\pi^{1/4}\omega(t)^{3/2}} = 0. \label{wd}
\end{eqnarray}

\begin{figure}[htpb]
\centering
\includegraphics[width=\columnwidth]{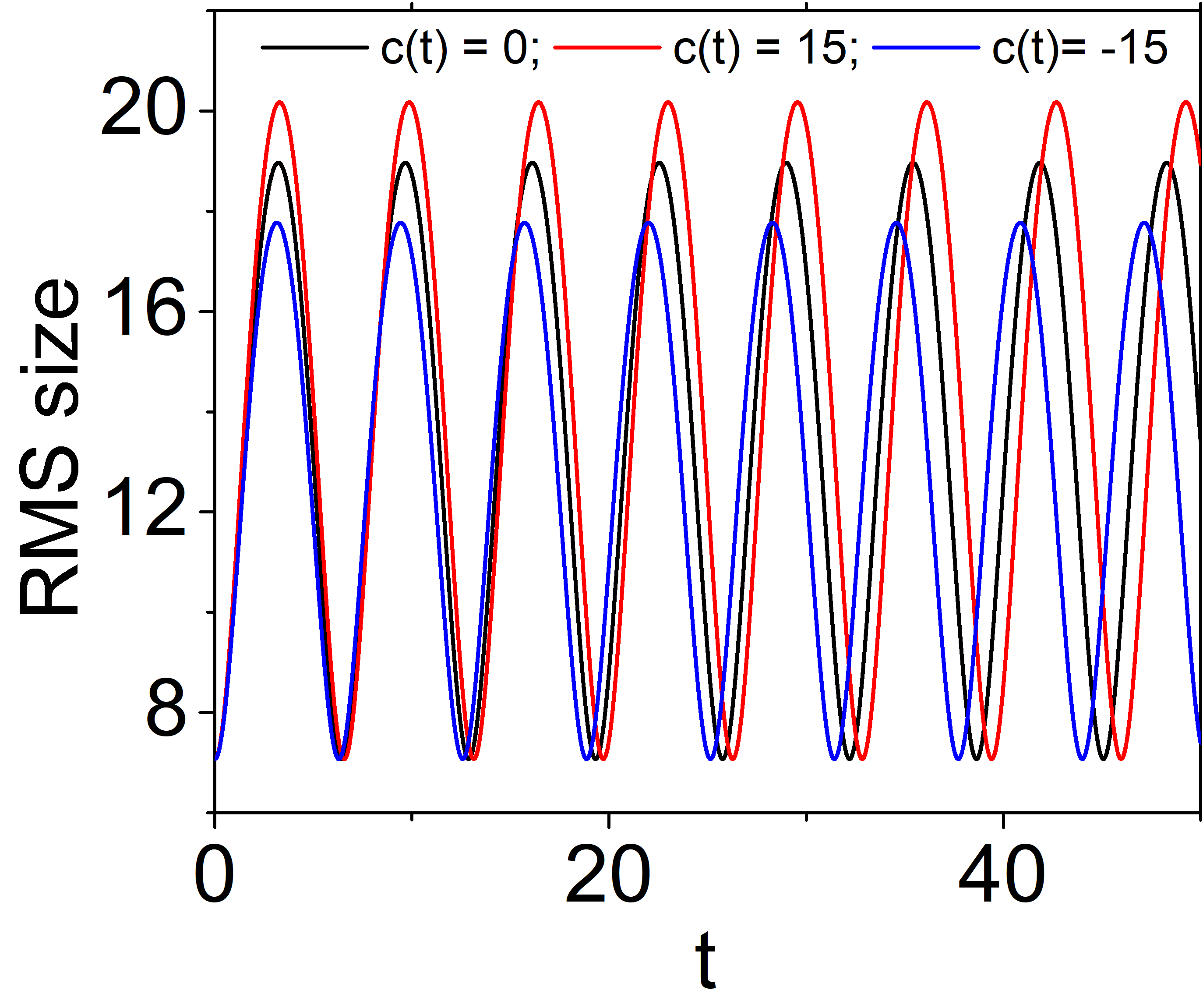}
\caption{\label{fig1} RMS size of the QDs oscillation with the time for $C(t) =$ 15 (red), 0 (black), and -15 (blue). Oscillation becomes faster as $C(t)$ goes positive to negative.}
\end{figure}

Equation (\ref{wd}) is a second-order non-linear differential equation governing the temporal evolution of the condensate width. The dynamics are controlled primarily by the harmonic trap strength $\gamma$ and the modulated spike amplitude $C(t)$. The last term encapsulates the influence of the LHY correction characteristic of the droplet regime.

To investigate the frequency response of the system, we perform a Fast Fourier Transform (FFT) of the time-dependent width obtained from Eq.~(\ref{wd}). Using varying $C(t)$, we explore different regimes of effective confinement. For positive $C(t)$, the composite potential mimics a double-well structure, while for negative $C(t)$, it produces a dimple in the center of a harmonic trap. These configurations enable precise control over droplet oscillations and stability.

Our analytical results are corroborated by numerical simulations of the eGPE using a split-step Fourier method, showing excellent agreement. These findings highlight the effectiveness of temporal modulations in manipulating quantum droplet dynamics and may serve as a basis for experimental exploration of driven BEC droplets.

\begin{figure}[htpb]
\centering
\includegraphics[width=\columnwidth]{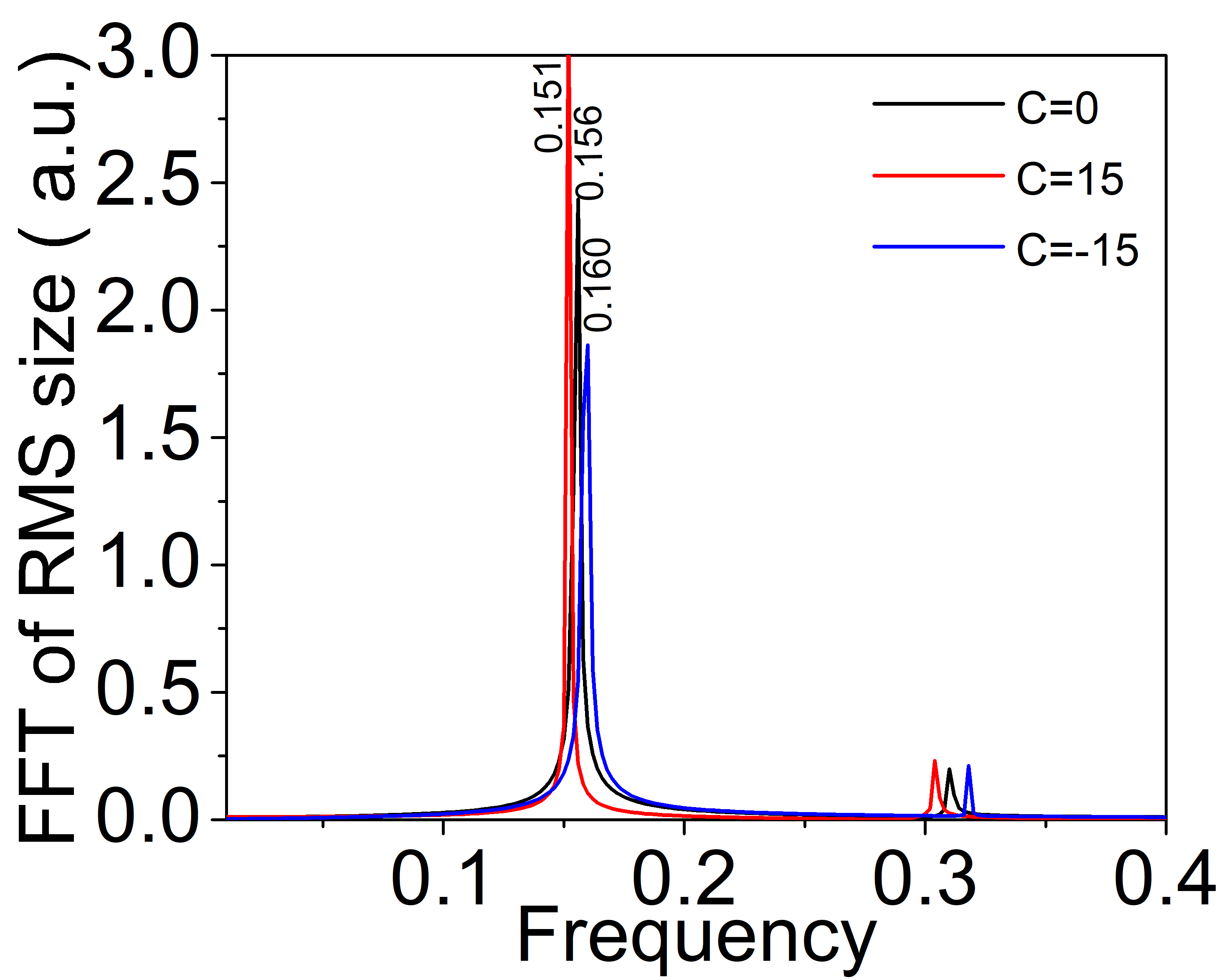}
\caption{\label{fig2} This is the frequency spectrum corresponding to the RMS size of the QDs oscillation with the time for $C(t) =$ 15 (red), 0 (black), and -15 (blue). Frequency is shifted towards right as $C(t)$ goes positive to negative.}
\end{figure}

\section{Dynamics of Quantum Droplets under Time-Dependent Gaussian Potentials}
In this section, we systematically investigate the dynamical response of 1D QDs under three experimentally relevant scenarios involving time-dependent external potentials. The central focus is on the influence of a localized Gaussian spike potential superimposed on a harmonic trap. The dynamics are probed through the time evolution of the RMS size of the QD, defined as
\begin{equation}
x_{\mathrm{rms}}(t) = \sqrt{\langle x^2 \rangle - \langle x \rangle^2},
\end{equation}
and further analyzed via FFT to extract the dominant oscillation frequencies.

We consider the following three configurations for the Gaussian spike amplitude \( C(t) \):
\begin{itemize}
    \item \textbf{Case A}: A constant positive spike strength, \( C(t) = +15 \), representing a repulsive Gaussian barrier (double-well-like potential).
    \item \textbf{Case B}: A constant negative spike strength, \( C(t) = -15 \), forming an attractive central dimple on top of the harmonic trap.
    \item \textbf{Case C}: A time-periodically modulated spike amplitude of the form 
    \begin{equation}
    C(t) = C_0 \left(1 + \alpha \cos(q t)\right),
    \end{equation}
    where \( C_0 = 15 \), \( \alpha = 0.5 \), and \( q = 0.05 \), representing a sinusoidally modulated potential depth.
\end{itemize}

For all cases, we numerically solve the 1D eGPE (Eq.~\ref{eq:QD1}) with fixed parameters: \( N_A = 4000 \), \( \sigma = 10.5 \), and \( \gamma = 0.55 \). We first analyze the unmodulated cases (A) and (B), comparing the dynamics with and without the central spike.

\begin{figure}[t]
\centering
\includegraphics[width=\columnwidth]{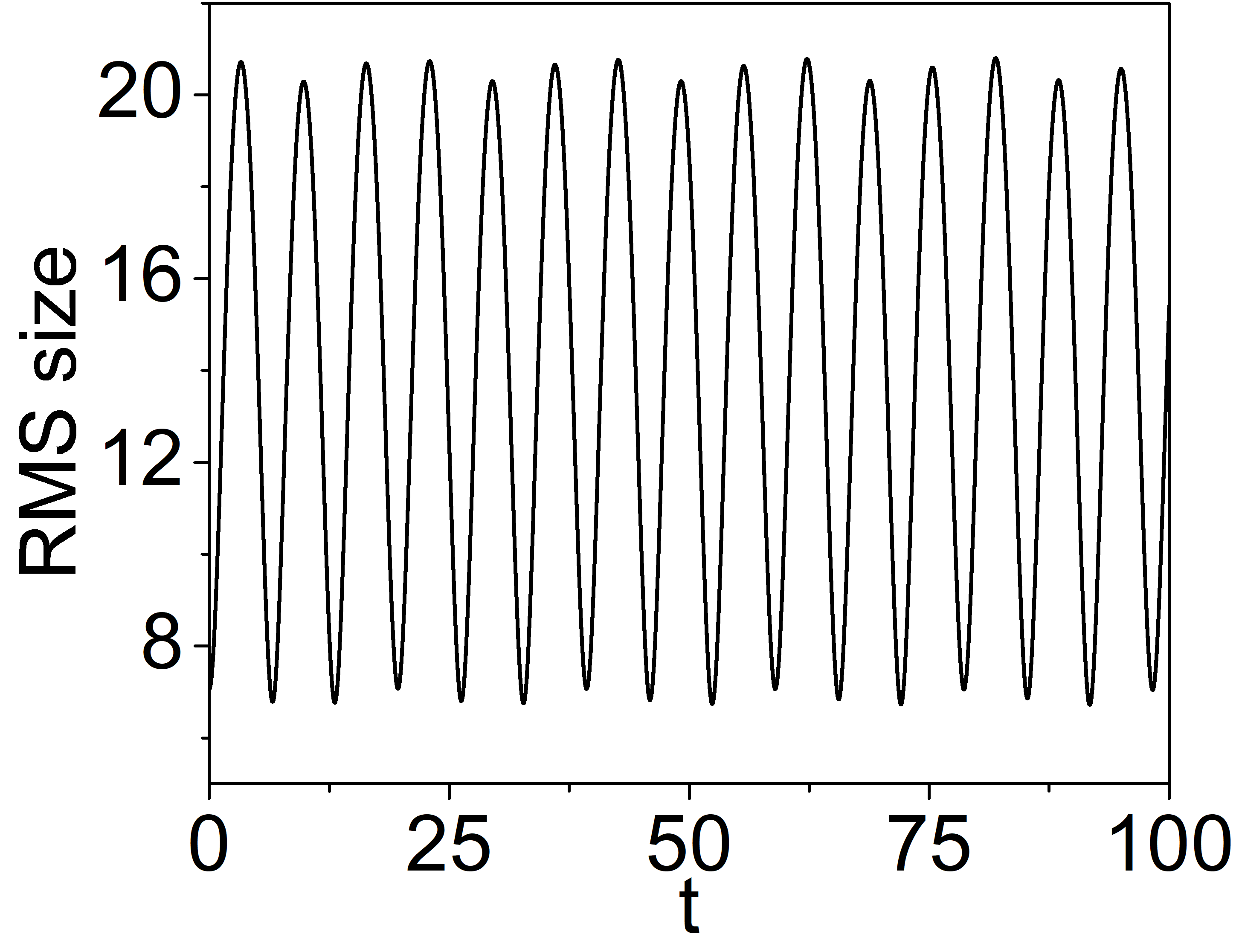}
\caption{\label{fig3} RMS size of the QDs oscillation with the time for $C(t) = C_0(1 +\alpha~ cos(qt))$. Here, $C_0=15$, $\alpha=0.5$ and $q=0.05$.}
\end{figure}

\subsection{Stationary Spike Potentials: Cases (A) and (B)}

Figure~\ref{fig1} presents the time evolution of the RMS size for \( C(t) = 15 \), \( 0 \), and \( -15 \). In all cases, the dynamics are dominantly oscillatory due to the underlying harmonic confinement. However, the presence of the Gaussian spike introduces subtle modifications in oscillation frequency and amplitude. Specifically, a repulsive barrier (positive \( C(t) \)) compresses the droplet, resulting in a higher frequency, while an attractive dimple (negative \( C(t) \)) allows for wider spreading and hence a lower oscillation frequency.

The corresponding frequency spectra, shown in Fig.~\ref{fig2}, reveal fundamental peaks at \( f = 0.151 \), \( 0.156 \), and \( 0.160 \) for \( C(t) = 15 \), \( 0 \), and \( -15 \), respectively. The presence of harmonics in the FFT spectra indicates nonlinear oscillatory behavior, which is consistent with the nonlinear nature of the governing dynamics. Moreover, the shift in the fundamental frequency is observed to vary linearly with the spike amplitude \( C(t) \), as quantified in Fig.~\ref{fig4a}, where we fit the frequency variation with:
\begin{equation}
f(C) = 0.1562 - 2.38 \times 10^{-4} C.
\end{equation}
This linear dependence implies that precise control over the Gaussian spike strength offers a tunable method to engineer desired oscillation frequencies in the droplet dynamics.

\subsection{Time-Modulated Spike Potential: Case (C)}

We now explore the effect of a time-periodically modulated spike amplitude as described by Eq.~(7). This scenario introduces an explicit time-dependent driving frequency \( q = 0.05 \) into the system, effectively realizing a parametrically driven nonlinear oscillator.

\begin{figure}[t]
\centering
\includegraphics[width=\columnwidth]{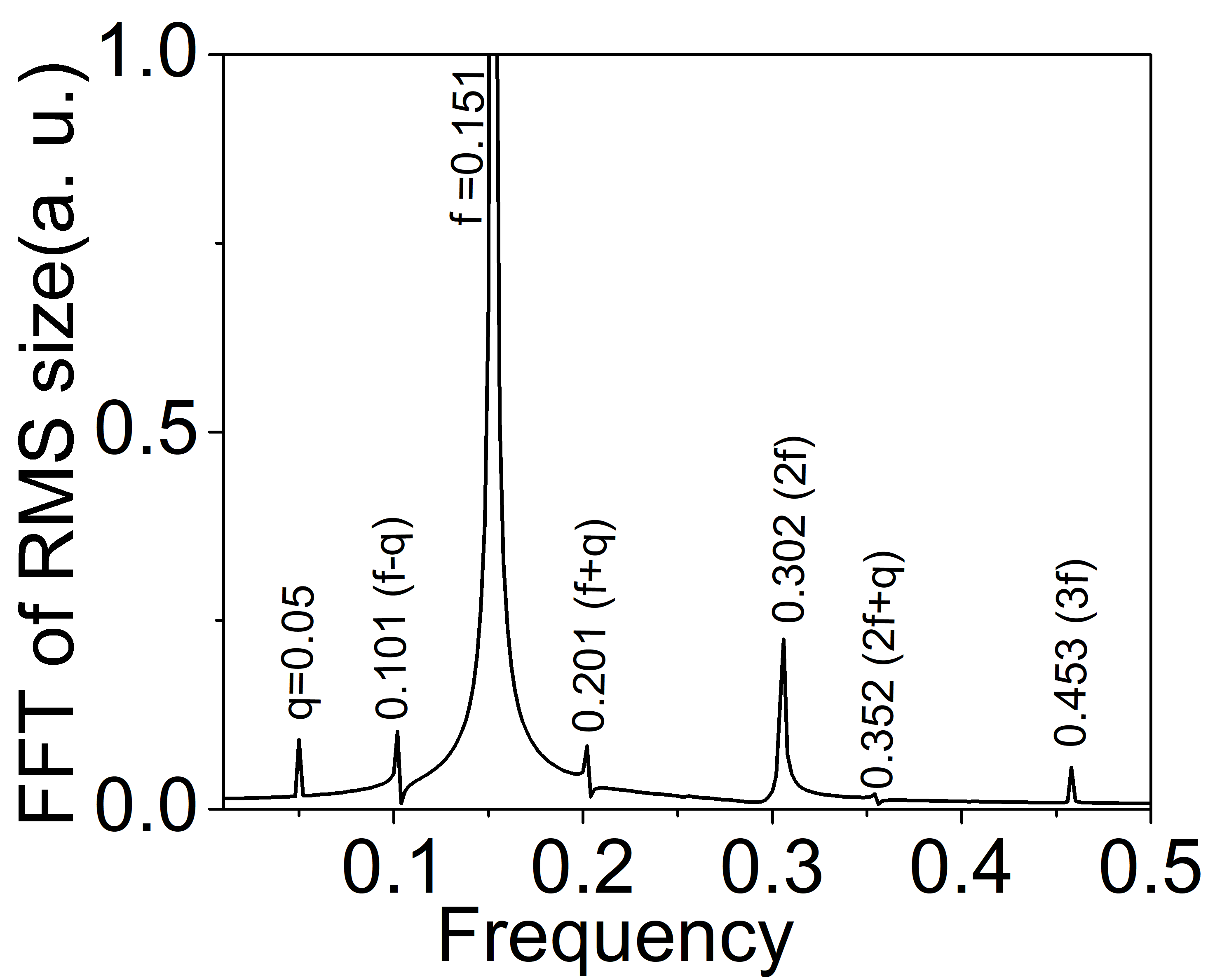}
\caption{\label{fig4} Frequency spectrum corresponding to the RMS size of the QDs oscillation with the time for $C(t) = C_0(1 +\alpha~ cos(qt))$. Here, $\gamma=0.55$ $C_0=15$, $\alpha=0.5$ and $q=0.05$. All the observed frequency spikes are labeled as, the modulation frequency $q= 0.05$ with which spike potential is chirping and fundamental frequency $f=0.151$, along with that some other combinations of frequencies $0.101 (f-q)$, $0.201(f+q)$ and $0.352(2f+q)$, higher harmonics of fundamental frequency$(f)$, $2f=0.302$ and $3f=0.453$ are present.}
\end{figure}

Figure~\ref{fig3} shows the RMS size evolution under periodic modulation. Unlike the quasi-sinusoidal oscillations in cases (A) and (B), the dynamics here display a complex beating pattern, indicative of multiple interacting frequency components. The FFT spectrum in Fig.~\ref{fig4} confirms the presence of the intrinsic droplet mode (\( f \approx 0.151 \)) as well as the external driving frequency \( q = 0.05 \). Additionally, sidebands at combination frequencies \( f \pm q \) (i.e., \( 0.101 \) and \( 0.201 \)) and higher harmonics are clearly resolved.

This spectral signature is a hallmark of nonlinear frequency mixing and modulation instability, phenomena that arise from the intrinsic nonlinearity of the droplet system. The observation of these mixed-frequency modes validates the use of QDs as nonlinear oscillators, where external modulations can induce controlled frequency splitting and spectral broadening.

To complete the dynamical picture, Fig.~\ref{fig4b} presents the spatiotemporal evolution of the QD density \( |\psi(x,t)|^2 \), further confirming the modulation-induced perturbations superimposed on the otherwise periodic breathing modes. Additionally, we investigate the impact of tuning of $\gamma$ tuning from $0.55 \xrightarrow{0.45} $ in figures (4) and (5). It apparent from the figure that the increase in $\gamma$ leads to enhancement in the magnitude of generated fundamental and mixed frequencies in the QDs.

\begin{figure}[t]
\centering
\includegraphics[width=\columnwidth]{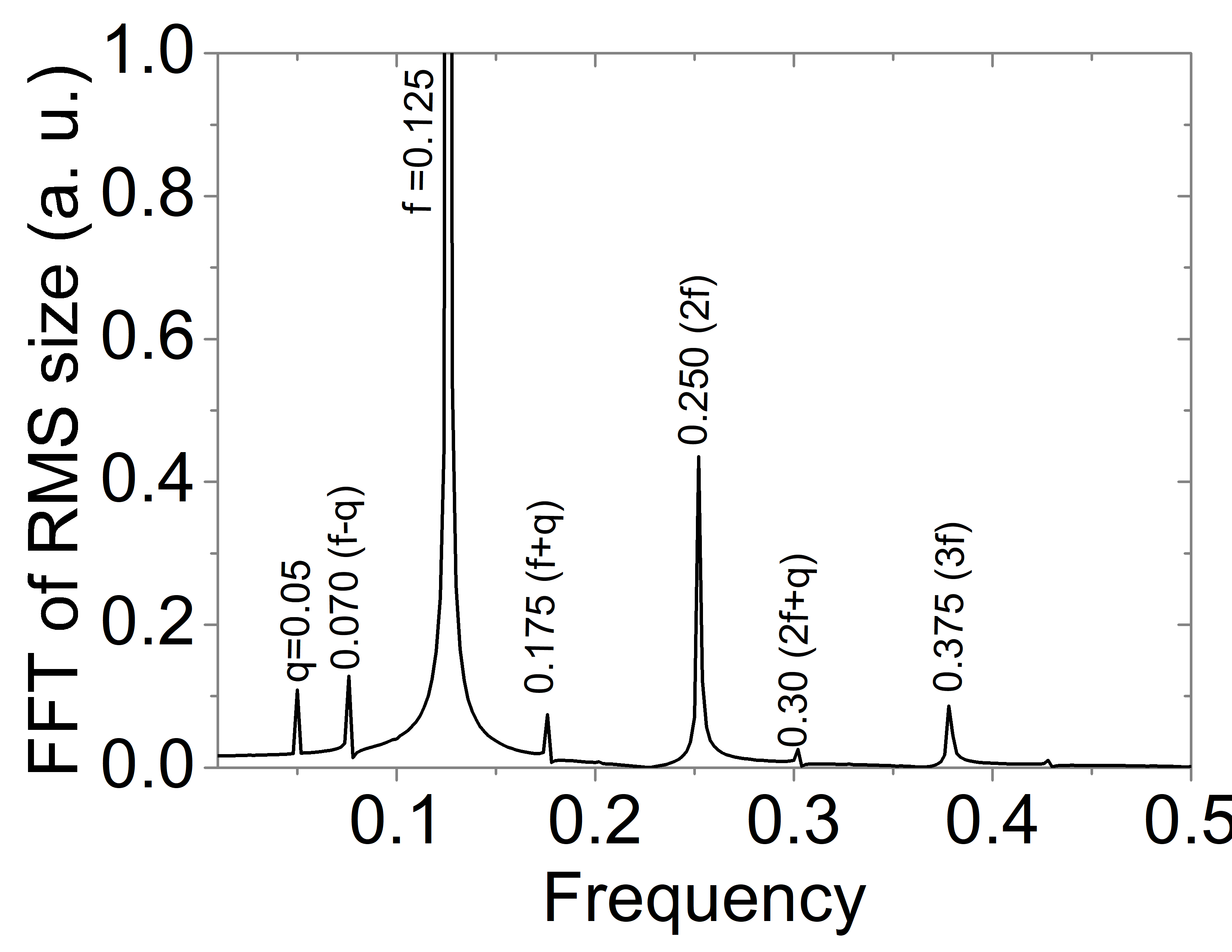}
\caption{\label{fig4a0} Frequency spectrum corresponding to the RMS size of the QDs oscillation with the time for $C(t) = C_0(1 +\alpha~ cos(qt))$. Here, $\gamma=0.45$ $C_0=15$, $\alpha=0.5$ and $q=0.05$. All the observed frequency spikes are labeled as, the modulation frequency $q= 0.05$ with which spike potential is chirping and fundamental frequency $f=0.125$, along with that some other combinations of frequencies $0.070 (f-q)$, $0.175(f+q)$ and $0.30(2f+q)$, higher harmonics of fundamental frequency$(f)$, $2f=0.25$ and $3f=0.375$ are present.}
\end{figure}

\begin{figure}[t]
\centering
\includegraphics[width=\columnwidth]{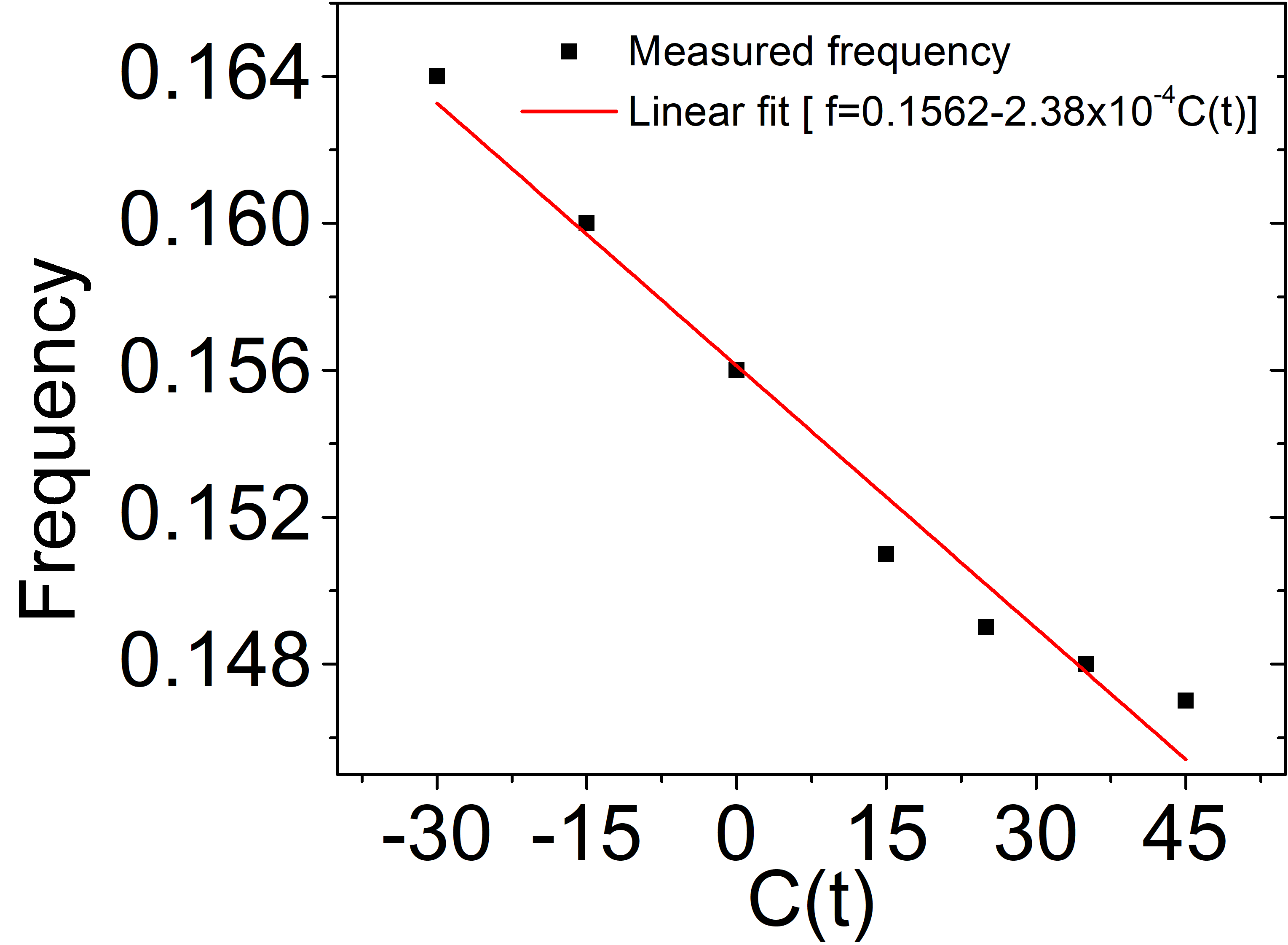}
\caption{\label{fig4a} Fundamental frequency variations with the Gs amplitude. A linear relation between $f$ and $C(t)$ is $f=0.1562 -2.38\times10^{-4}C(t)$ is derived. Frequency decreases with the in creasing $C(t)$ and effective harmonic trap is 0.1562. }
\end{figure}

\begin{figure}[t]
\centering
\includegraphics[width=\columnwidth]{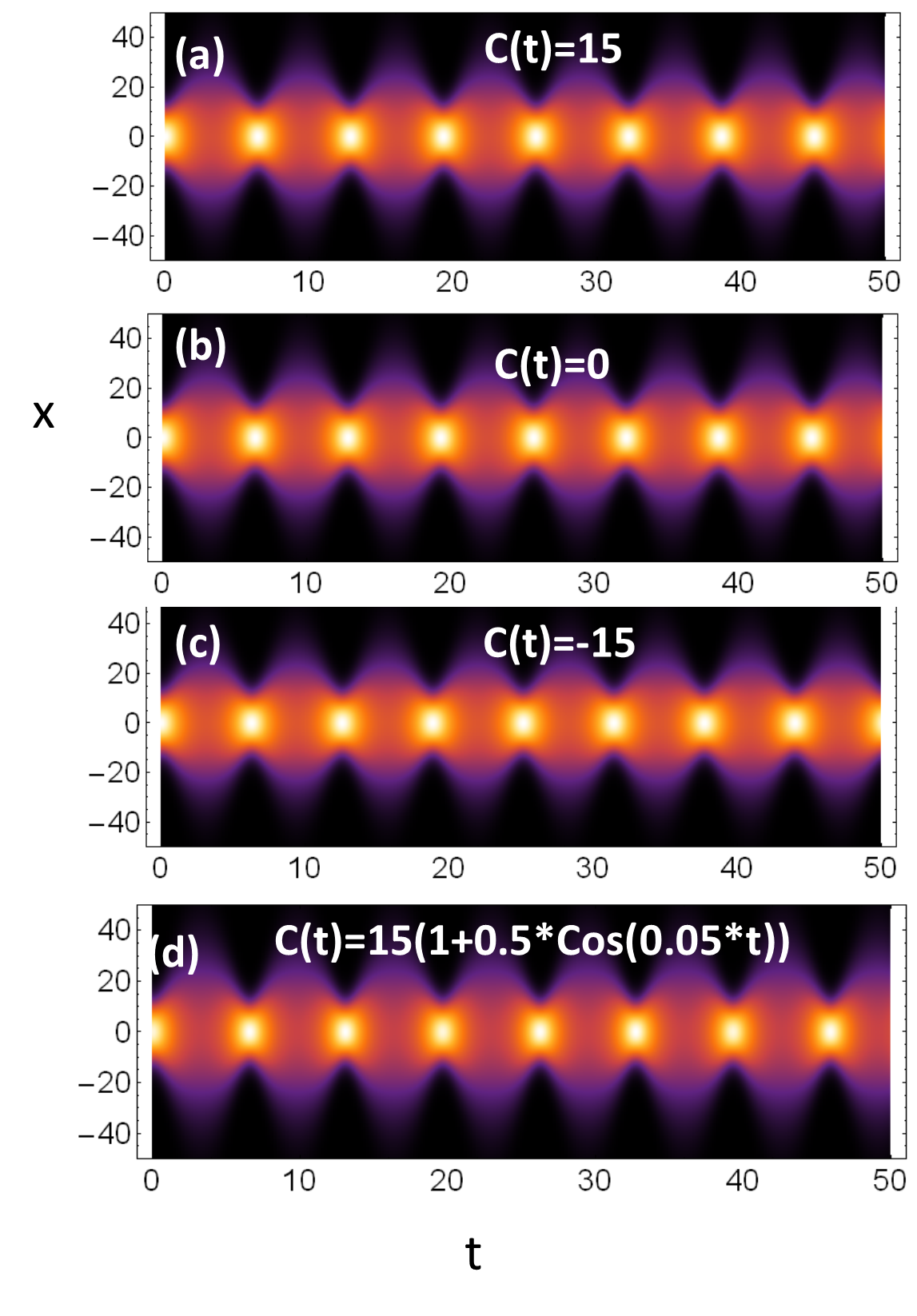}
\caption{\label{fig4b} In this figure density of the QDs variations with the time is presented for $c(t)= 15,~0,~-15$ and $15(1+0.5~cos(0.5t))$ from (a) to (D), respectively.}
\end{figure}

\section{Numerical Simulation and Phase-Space Analysis} In this section, we present a detailed numerical investigation of the quantum droplet (QD) dynamics under all three scenarios discussed previously: (A) constant repulsive Gaussian spike, (B) constant attractive Gaussian spike, and (C) periodically modulated Gaussian spike amplitude. The governing extended Gross-Pitaevskii equation (Eq.~\ref{eq:QD1}) is solved numerically using the Split-Step Fourier Transform (SSFT) method implemented in MATLAB. The numerical simulations are found to be in strong agreement with the results obtained from the analytical variational method, validating the robustness of the analytical model.

For the simulations, we used a spatial grid spacing \( dx = 0.0078 \), a temporal step size \( dt = 0.05 \), and evolved the system for 2500 time steps. The system parameters and potential configurations were kept identical to those used in the variational analysis to enable a direct comparison.

\begin{figure}[t]
\centering
\includegraphics[width=\linewidth]{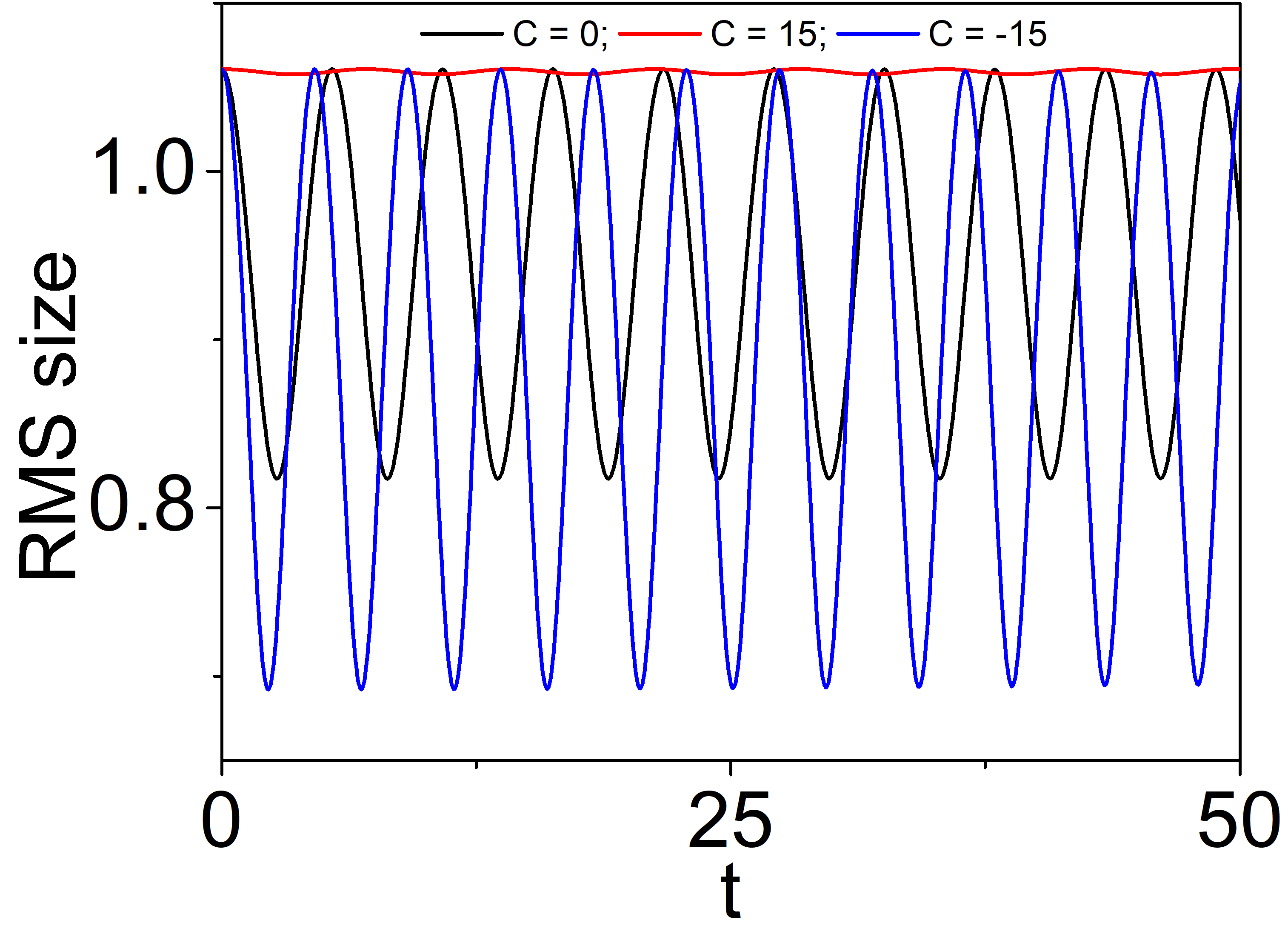}
\caption{\label{fig5} RMS size of the QDs oscillation with the time for $C(t) =$ 15 (red), 0 (black), and -15 (blue).}
\end{figure}

\begin{figure}[t]
\centering
\includegraphics[width=\linewidth]{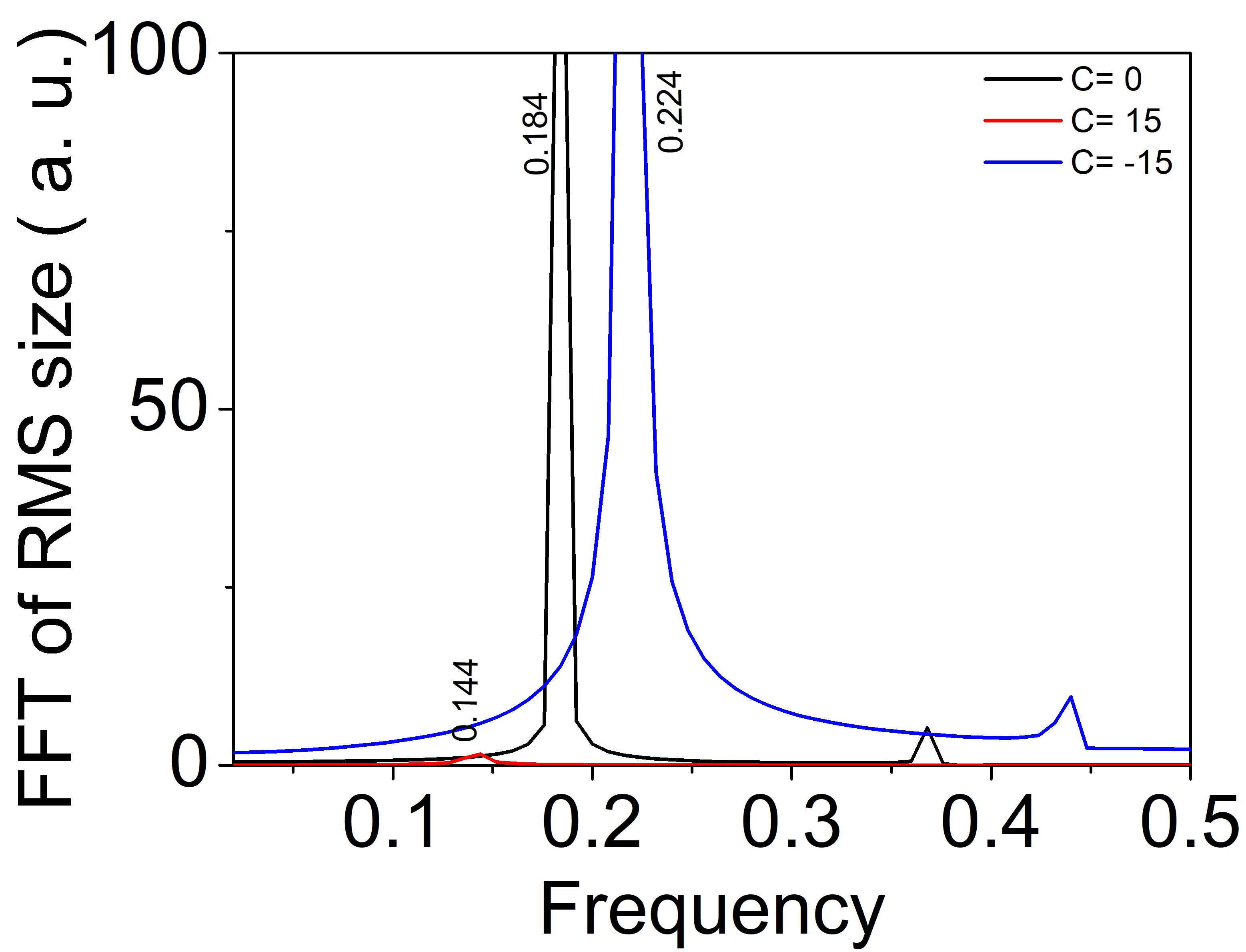}
\caption{\label{fig6} This is the frequency spectrum corresponding to the RMS size of the QDs oscillation with the time for $C(t) =$ 15 (red), 0 (black), and -15 (blue). Frequency is shifted towards right as $C(t)$ goes positive to negative.}
\end{figure}

Figures~\ref{fig5} and \ref{fig6} show the time evolution of the RMS size and the corresponding frequency spectra for cases (A) and (B), respectively. The oscillatory behavior of the RMS size and the presence of distinct peaks in the FFT spectra corroborate the analytical predictions. The fundamental frequencies observed in the numerical spectra are \( f = 0.144 \), \( 0.184 \), and \( 0.224 \) for \( C(t) = 15 \), \( 0 \), and \( -15 \), respectively. While the trends and spectral positions are consistent with the variational results, slight discrepancies in amplitude arise due to the approximations inherent in the analytical method.

\begin{figure}[t]
\centering
\includegraphics[width=\linewidth]{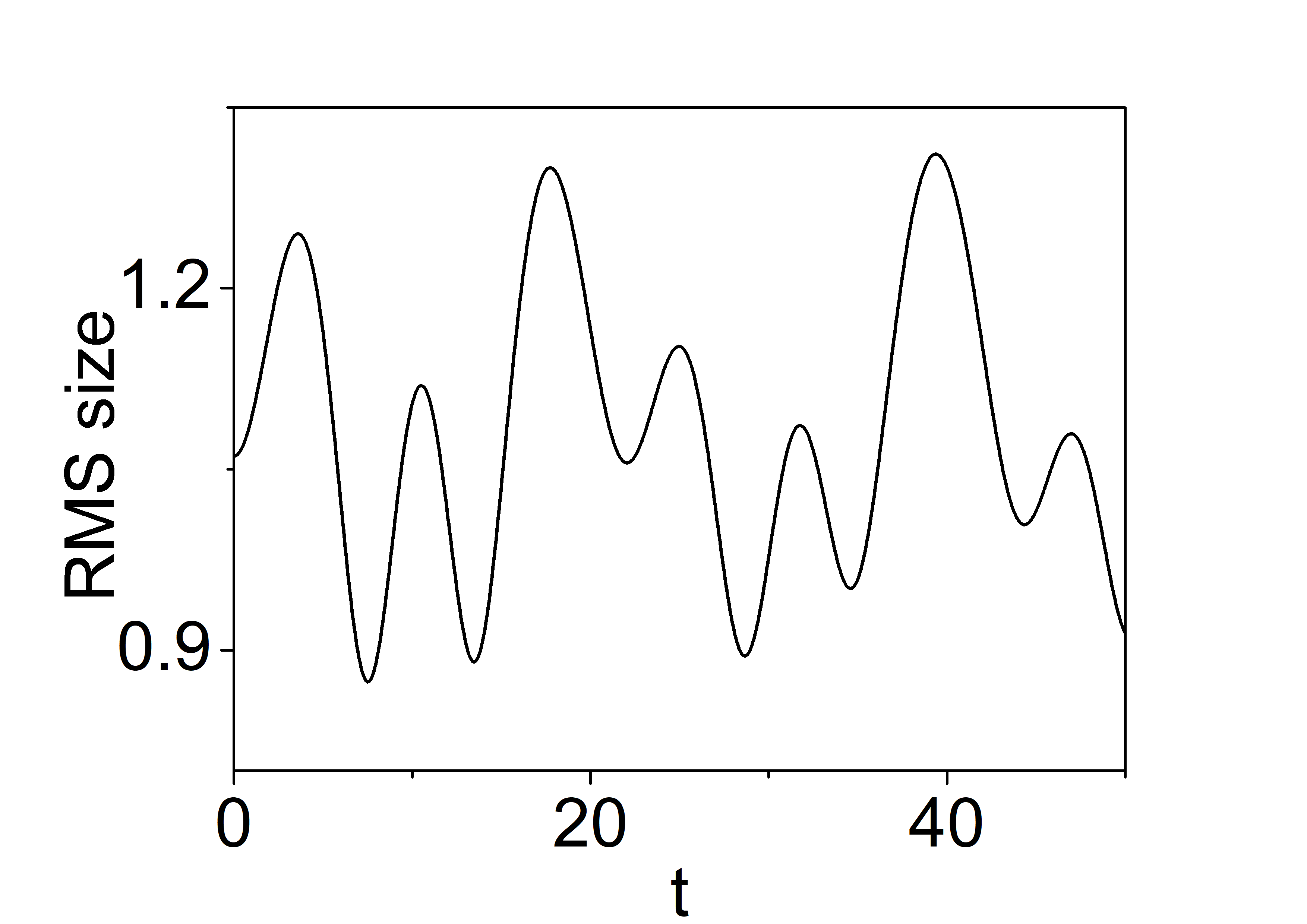}
\caption{\label{fig7} RMS size of the QDs oscillation with the time for $C(t) = C_0(1 +\alpha~ cos(qt))$. Here, $C_0=15$, $\alpha=0.5$ and $q=0.05$.}
\end{figure}

\begin{figure}[t]
\centering
\includegraphics[width=\linewidth]{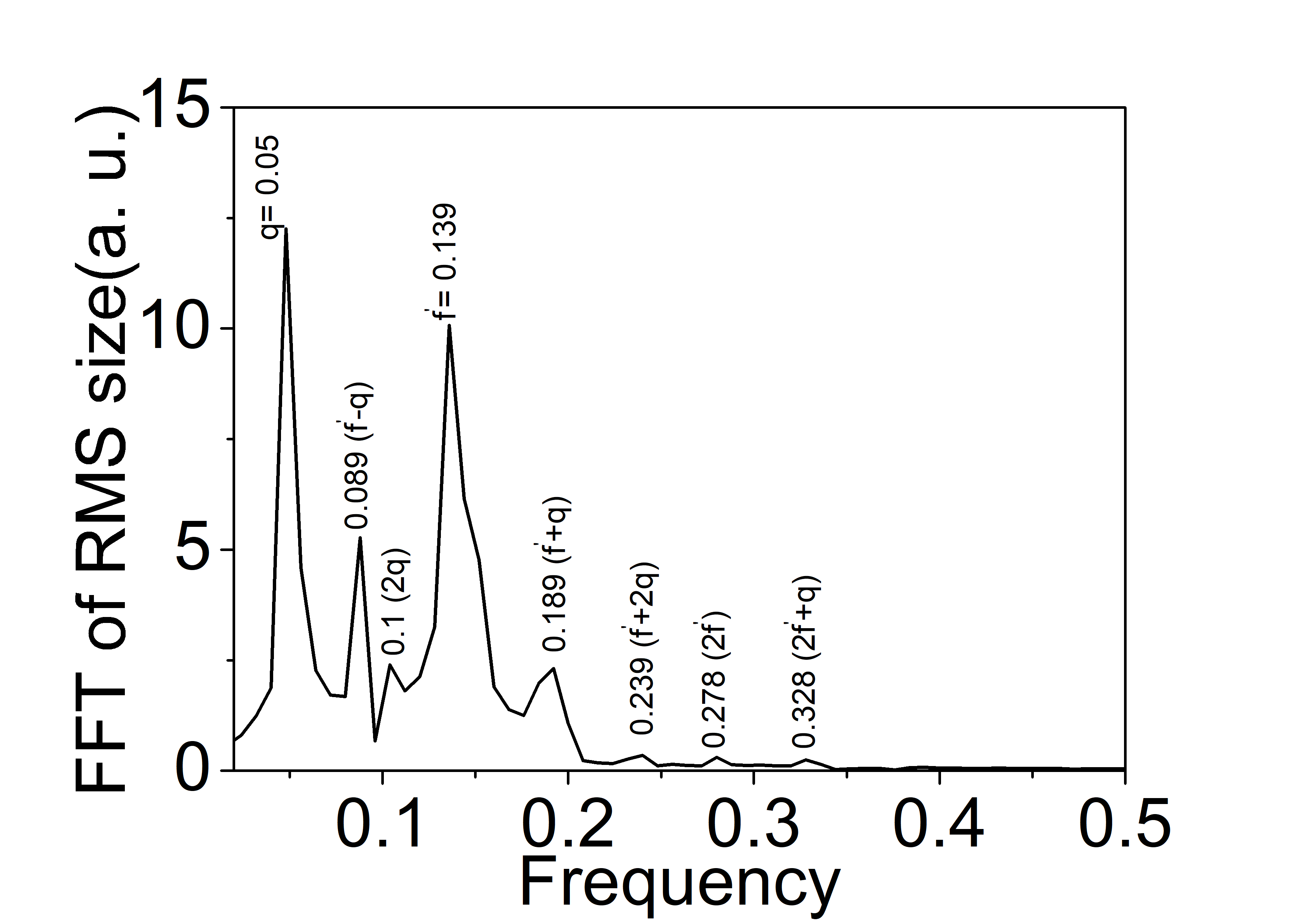}
\caption{\label{fig8} This is the frequency spectrum of RMS size of the QDs oscillation with the time for $C(t) = C_0(1 +\alpha~ cos(qt))$. Here, $C_0=15$, $\alpha=0.5$ and $q=0.05$. All the observed frequencies are labeled in the frequency spectrum as: $q = 0.05$ (chirping frequency), $0.089 (f^{\prime}-q)$, $0.1 (2q)$, $f^{\prime}=0.139$ (fundamental frequency), $0.189 (f^{\prime}+q)$, $0.239 (f^{\prime}+2q)$, $0.278 (2f^{\prime})$ and $0.328(2f^{\prime}+q)$.}
\end{figure}

For case (C), where the spike amplitude is modulated periodically in time as
\[
C(t) = C_0 \left(1 + \alpha \cos(q t)\right),
\]
with \( C_0 = 15 \), \( \alpha = 0.5 \), and \( q = 0.05 \), the RMS size dynamics and the corresponding FFT spectrum are depicted in Figs.~\ref{fig7} and \ref{fig8}, respectively. The numerical results closely mirror the analytical findings, showing rich spectral features due to nonlinear frequency mixing. The observed frequency components include the modulation frequency \( q = 0.05 \), the fundamental droplet oscillation frequency \( f' = 0.139 \), and their combinations such as \( f' \pm q \), \( 2f' \), \( 2q \), and \( 2f' + q \). Specifically, the identified peaks are at: \( 0.089 = f' - q \), \( 0.1 = 2q \), \( 0.139 = f' \), \( 0.189 = f' + q \), \( 0.239 = f' + 2q \), \( 0.278 = 2f' \), and \( 0.328 = 2f' + q \). These spectral features underscore the nonlinear nature of the QD dynamics, where higher harmonics and frequency mixing arise due to the interplay between the intrinsic droplet mode and the external temporal modulation.

\section{\textit{Wigner Phase-Space Distribution}}

To further probe the quantum dynamics and phase coherence properties of the droplet, we analyze the system using the Wigner quasi-probability distribution function. This distribution provides a powerful tool for visualizing quantum states in phase space and is defined as:
\begin{equation}
W(x,p) = \frac{1}{\pi\hbar} \int_{-\infty}^{\infty} \psi^*(x + x_0, t) \psi(x - x_0, t) \, e^{2 i x_0 p / \hbar} \, dx_0.
\end{equation}
The Wigner function combines both position and momentum (phase) information and is especially sensitive to quantum interference effects. For instance, nonclassical states such as Schrödinger-cat states manifest as interference ripples in this representation.

\begin{figure}[t]
\centering
\includegraphics[width=\linewidth]{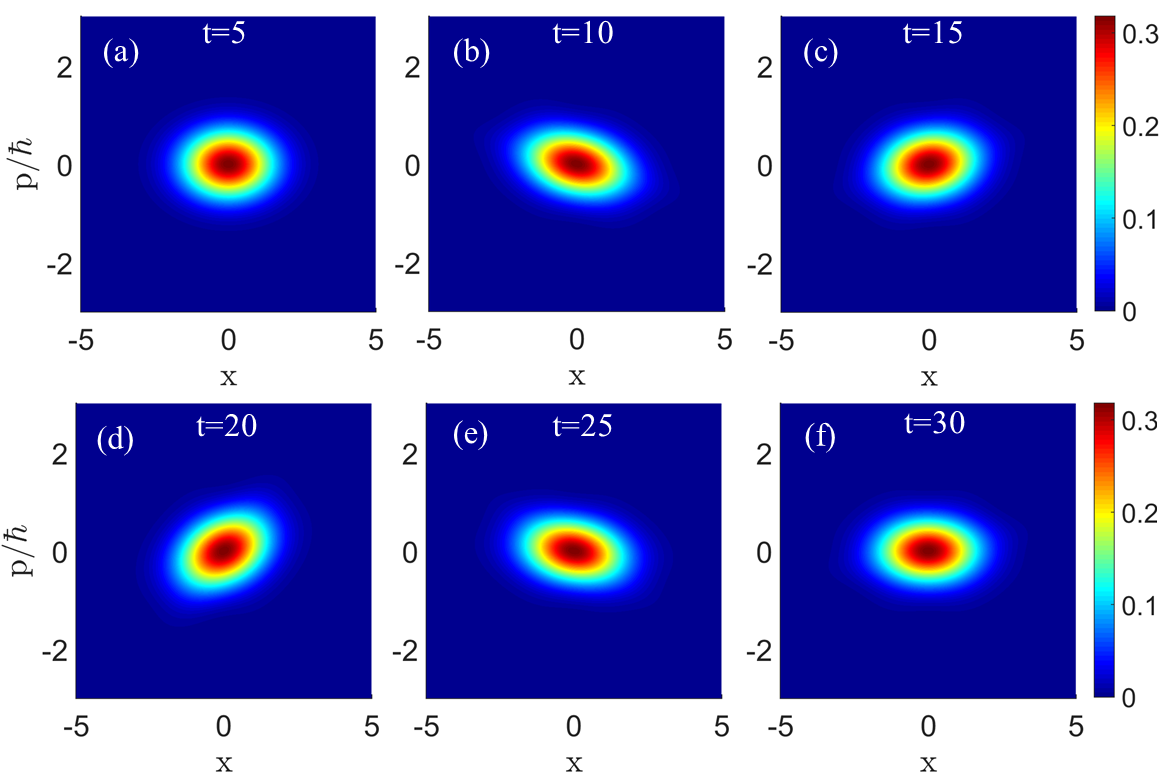}
\caption{\label{fig9} The Wigner phase-space distribution is depicted here in the figure for $C(t) = C_0(1 +\alpha~ cos(qt))$. Here, $C_0=15$, $\alpha=0.5$ and $q=0.05$ at different time.}
\end{figure}

Figure~\ref{fig9} displays snapshots of the Wigner phase-space distribution at times \( t = 5, 10, 15, 20, 25, \) and \( 30 \) for case (C). Although the real-space density profiles of the QD remain similar across these times, the Wigner distributions clearly show rotational patterns in phase space, indicating evolving phase structures. This highlights an important insight: even when the density appears stationary or quasi-periodic, subtle changes in the phase of the condensate wavefunction can be captured through the Wigner function.

Such phase-sensitive visualization reveals internal mode excitations and weak interactions that may be invisible in traditional density-based diagnostics. Therefore, Wigner phase-space analysis serves as a complementary diagnostic tool for understanding the full quantum dynamics of nonlinear systems like quantum droplets, especially in scenarios involving time-dependent potentials or external modulations. 

\section{Conclusion} We have investigated the dynamics of one-dimensional QDs by solving the 1D extended Gross-Pitaevskii equation under a harmonic confinement augmented by a static or time-dependent Gs potential. Using both a variational analytical method and full numerical simulations, we analyzed the evolution of the droplet's RMS size, frequency response, and phase-space structure. Our results show that while the harmonic trap sets the dominant confinement scale, the Gs potential offers precise control over QD dynamics. Tuning the Gs amplitude from positive (repulsive) to negative (attractive) systematically increases the frequency of size oscillations, allowing fine control over excitation spectra. When the Gs amplitude is periodically modulated, the QD exhibits nonlinear dynamical features, including higher harmonics and frequency mixing, characteristic of driven many-body systems. A Wigner phase-space analysis reveals coherent rotational evolution, highlighting phase dynamics that are not evident in real-space density alone. This demonstrates the potential of phase-space diagnostics for probing coherence and internal modes in QDs.
These findings open new possibilities for controlling QD excitations in ultracold atomic systems. The demonstrated tunability and nonlinear response provide a foundation for applications in quantum simulation of driven systems, atomtronic devices with frequency-selective control, and the study of coherence and phase transitions in low-dimensional quantum matter.

\section{Appendix A}
The underlying dynamics can be solved analytically by using Lagrangian variational (LV) approach under some approximations \cite{Otajonov, Chen, Saha}. We start by considering the following ansatz solution,
\begin{eqnarray}
\psi(x,t)=A(t)\exp\Big[-\frac{x^2}{2\omega(t)^2}+i\beta(t)x^2+i\phi(t)\Big],\label{ansatz}
\end{eqnarray}
where $\omega(t)$ is the width, $\beta(t)$ is the chirping parameter and $\phi(t)$ is a absolute phase.
The Lagrangian density corresponding to the above ansatz is written as,

\begin{eqnarray}
l_{g}=\frac{i}{2}\Big(\psi(x,t)\frac{\partial\psi^{*}(x,t)}{\partial t}-\psi^{*}(x,t)\frac{\partial\psi(x,t)}{\partial t}\Big)\nonumber\\
+\frac{1}{2}\Big|\frac{\partial\psi(x,t)}{\partial x}\Big|^{2}+\frac{1}{2}\alpha(t)^2x^2|\psi(x,t)|^{2}\nonumber\\+ C(t)\exp\Big(\frac{-x^2}{2\sigma^2}\Big)|\psi(x,t)|^2
+\frac{1}{2}|\psi(x,t)|^4-\frac{1}{2}\psi(x,t)|^\frac{3}{2}\nonumber\\
\end{eqnarray}

\begin{eqnarray}
 l_ 1= \frac{i}{2}\Big(\psi(x,t)\frac{\partial\psi^{*}(x,t)}{\partial t}-\psi^{*}(x,t)\frac{\partial\psi(x,t)}{\partial t}\Big)\nonumber\\
 =\exp\Big(-\frac{x^2}{\omega(t)^2}\Big)A(t)^2 \Big(x^2\beta^\prime(t)+\phi^\prime(t)
\Big )
 \end{eqnarray}
 
 \begin{eqnarray}
L_1=\int_{-\infty}^{\infty}\exp\Big(-\frac{x^2}{\omega(t)^2}\Big)A(t)^2 \Big(x^2\beta^\prime(t)+\phi^\prime(t)\Big )dx\nonumber\\
=\frac{\sqrt{\pi}A(t)^2(\omega(t)^2\beta^\prime(t)+2\phi^\prime(t))}{2\sqrt\frac{1}{\omega(t)^2}}
 \end{eqnarray}
 
\begin{eqnarray}
l_2= \frac{1}{2}\Big|\frac{\partial\psi(x,t)}{\partial x}\Big|^{2}
&=\frac{\exp\Big(-\frac{x^2}{\omega(t)^2}x^2A(t)^2(1+4\beta(t)^2\omega(t)^4)\Big)}{2\omega(t)^4}
\nonumber\\
\end{eqnarray}

\begin{eqnarray}
L_2=\int_{-\infty}^{\infty}\frac{\exp\Big(-\frac{x^2}{\omega(t)^2}x^2A(t)^2(1+4\beta(t)^2\omega(t)^4)\Big)}{2\omega(t)^4}dx\nonumber\\ 
=\frac{1}{4}\sqrt{\pi}A(t)^2\sqrt{\frac{1}{\omega(t)^2}}(1+4\beta(t)^2\omega(t)^4)
\end{eqnarray}

\begin{eqnarray}
l_3=\frac{1}{2}\gamma^2x^2|\psi(x,t)|^{2}+C(t)\exp\Big(\frac{-x^2}{2\sigma^2}\Big)|\psi(x,t)|^2\nonumber\\
=\exp\Big(-\frac{x^2}{2\sigma^2}-\frac{x^2}{\omega(t)^2}\Big)A(t)^2C(t)+\nonumber\\
\frac{1}{2}\exp\Big(-\frac{x^2}{\omega(t)^2}\Big)x^2A(t)^2\gamma^2
\end{eqnarray}

\begin{eqnarray}
L_3=\int_{-\infty}^{\infty}\Big(\exp\Big(-\frac{x^2}{2\sigma^2}-\frac{x^2}{\omega(t)^2}\Big)A(t)^2C(t)+\nonumber\\
\frac{1}{2}\exp\Big(-\frac{x^2}{\omega(t)^2}\Big)x^2A(t)^2\gamma^2\Big)dx \nonumber\\
= \frac{\sqrt{2\pi}A(t)^2C(t)}{\sqrt{\frac{1}{\sigma^2}+\frac{2}{\omega(t)^2}}}+\frac{\sqrt{\pi}A(t)^2\gamma^2}{4(\frac{1}{\omega(t)^2})^\frac{3}{2}}
\end{eqnarray}

\begin{eqnarray}
l_4= \frac{1}{2}|\psi(x,t)|^4
= \frac{1}{2}\exp\Big(-\frac{2x^2}{\omega(t)^2}\Big)A(t)^4
\end{eqnarray}

\begin{eqnarray}
L_4=\int_{-\infty}^{\infty}\frac{1}{2}\exp\Big(-\frac{2x^2}{\omega(t)^2}\Big)A(t)^4dx
=\frac{\sqrt{\frac{\pi}{2}}A(t)^4}{2\sqrt{\frac{1}{\omega(t)^2}}}
\end{eqnarray}

\begin{eqnarray}
l_5= -\frac{1}{2}\psi(x,t)|^\frac{3}{2}
=-\frac{2}{3}\Big(\exp\Big(-\frac{x^2}{\omega(t)^2}\Big)A(t)^2\Big)^\frac{3}{2}
\end{eqnarray}

\begin{eqnarray}
L_5=\int_{-\infty}^{\infty}-\frac{2}{3}\Big(\exp\Big(-\frac{x^2}{\omega(t)^2}\Big)A(t)^2\Big)^\frac{3}{2}dx\nonumber\\
=-\frac{2\sqrt{\frac{2\pi}{3}}A(t)^3}{3\sqrt{\frac{1}{\omega(t)^2}}}
\end{eqnarray}

The effective total Lagrangian is given by
\begin{eqnarray}
L_{e\!f\!f}=\int_{-\infty}^{\infty}l_{g}dx\nonumber\\
=L_{1}+L_{2}+L_{3}+L_{4}+L_{5}\nonumber\\
=\frac{\sqrt{2\pi}A(t)^2C(t)}{\sqrt{\frac{1}{\sigma^2}+\frac{2}{\omega(t)^2}}}
+\frac{\sqrt{\pi}A(t)^2\gamma^2}{4\Big(\frac{1}{\omega(t)^2}\Big)^\frac{3}{2}}\nonumber\\
-\frac{2\sqrt{\frac{2\pi}{3}A(t)^3}}{3\sqrt{\frac{1}{\omega(t)^2}}}
+\frac{\sqrt{\frac{\pi}{2}}A(t)^4}{2\sqrt{\frac{1}{\omega(t)^2}}}\nonumber\\
+\frac{1}{4}\sqrt{\pi}A(t)^2\sqrt{\frac{1}{\omega(t)^2}(1+4\beta(t)^2\omega(t)^4)}\nonumber\\
+\frac{\sqrt{\pi}A(t)^2\Big(\omega(t)^2\beta^\prime(t)+2\phi^\prime(t)\Big)}{2\sqrt{\frac{1}{\omega(t)^2}}}
\end{eqnarray}

The corresponding Euler-Lagrange equation is given by
\begin{eqnarray}
\frac{d}{dt}\frac{\partial L_{e\!f\!f}}{\partial \dot{u}}=\frac{\partial L_{e\!f\!f}}{\partial u},\label{el}
\end{eqnarray}
where $u$ is one of the variational parameters, $\omega(t)$, $\beta(t)$ and $\phi(t)$. After solving Eq.(\ref{el})
for $u=\omega(t)$, $\beta(t)$ and $\phi(t)$, we obtain that the condensed width variation is derived as

\begin{eqnarray}
\omega ''(t)+\gamma^2 \omega (t)-\frac{1}{\omega (t)^3}-\frac{2 \sqrt{2} \sigma  c(t) \omega (t)}{\left(2 \sigma ^2+\omega (t)^2\right)^{3/2}}-\nonumber\\
\frac{\sqrt{\frac{N_{A}}{2 \pi }}}{\omega (t)^2}+\frac{2 \sqrt{6 N_{A}}}{9 \sqrt[4]{\pi } \omega (t)^{3/2}}=0\label{wd}
\end{eqnarray}

\end{document}